\documentclass[english]{sig-alternate-nofooter}

\usepackage{cite}
\usepackage{url}
\usepackage{caption}
\usepackage{subcaption}
\usepackage{graphicx}
\usepackage{color}
\usepackage{balance}

\newcommand{\ignore}[1]{}

\begin{document}

\title{Ten Blue Links on Mars}

\numberofauthors{4}
\author{
Charles L. A. Clarke, Gordon V. Cormack, Jimmy Lin, and Adam Roegiest\\[1ex]
\affaddr{David R. Cheriton School of Computer Science}\\
\affaddr{University of Waterloo, Ontario, Canada}\\[1ex]
\affaddr{claclark@gmail.com, \{gvcormac, jimmylin, aroegies\}@uwaterloo.ca}
}

\maketitle

\begin{abstract}
This paper explores a simple question:\ How would we provide a
high-quality search experience on Mars, where the fundamental physical
limit is speed-of-light propagation delays on the order of tens of
minutes? On Earth, users are accustomed to nearly instantaneous
response times from search engines. Is it possible to overcome
orders-of-magnitude longer latency to provide a tolerable
user experience on Mars? In this paper, we formulate the searching from Mars
problem as a tradeoff between ``effort'' (waiting for responses from
Earth) and ``data transfer'' (pre-fetching or caching data on
Mars). The contribution of our work is articulating this design space and
presenting two case studies that explore the effectiveness of baseline
techniques, using publicly available data from the TREC Total Recall
and Sessions Tracks. We intend for this research problem to be
aspirational and inspirational---even if one is not convinced by the
premise of Mars colonization, there are Earth-based scenarios such as
searching from a rural village in India that share similar
constraints, thus making the problem worthy of exploration and
attention from researchers.
\end{abstract}


\section{Introduction}

Search and other transactional web services
strive to minimize response times in order to provide a sense of
interactivity and to maintain user engagement.
Regardless of how efficiently we implement these services,
their response times are limited by roundtrip network latency,
which in turn is limited by technical and physical factors,
which include the speed of light.
For Earth-based users the physical limits imposed by the speed of light amount
to less than a second of delay,
even when a packet must bounce off a geosynchronous satellite.
Consider, however, the case of future colonists on Mars, who 
will be between 4 and 24 light-minutes away, depending on the relative positions of the two planets.\footnote{\url{http://blogs.esa.int/mex/2012/08/05/time-delay-between-mars-and-earth/}}
This paper explores a simple question:\ Is it possible to 
engineer around physical laws and provide
a tolerable search experience from Mars?

While Martian colonies may be a decade or more in the future,
plans are being actively developed,
with the public support of luminaries such as space entrepreneur
Elon Musk~\cite{space} and Edwin ``Buzz'' Aldrin,
one of the first two people to walk on the Moon~\cite{buzz}.
Both Mars to Stay\footnote{\url{http://www.marstostay.com/}}
and Mars One\footnote{\url{http://www.mars-one.com/}}
propose permanent settlements,
with colonists potentially living out the remainder of their lives on Mars.
While the idea of permanent settlements may seem like science fiction to some,
there are substantial cost savings from permanent colonization,
as opposed to a traditional Apollo-style there-and-back-again mission,
since fuel and other resources for immediate return will not be required.
Permanent colonists can conduct more science, over much longer periods,
greatly increasing the benefits accrued from the mission.

Current planning assumes that colonists will simply have to tolerate
communication delays, limiting their ability to use the Internet.
Mars One planners assume communication will be limited to email,
video messages, and the like.
For other services, they currently assume:

\begin{quote}
\small The astronauts can use the Internet,
but can only surf ``real time'' on a number of websites that are downloaded
from Earth on the Mars habitat webserver.
Every astronaut will have access to his favorite websites that way.
Visiting other websites will be a bit impractical because of the
delay.\footnote{\url{http://www.mars-one.com/faq/technology/how-does-the-mars-base-communicate-with-earth}} 
\end{quote}

While in the short term our colonists will tolerate whatever is necessary
for the success of the mission,
a long term separation from digital life on earth need not be one of them.
Searching, surfing, and shopping should be as easy from
Mars as it is from Marseille.


The primary contribution of this work is an articulation of the
design space of how we might engineer search from Mars. We model the
problem as a tradeoff between ``effort'' (waiting for responses from
Earth) and ``data transfer'' (pre-fetching or caching data on
Mars). We flesh out our design by considering two concrete tasks using
publicly available data: In the first task, we build on a previous
short paper~\cite{Clarke_etal_ICTIR2016} and explore high-recall retrieval (such as
conducting a scientific survey)\ using data from the TREC Total Recall
Track. In the second task, we simulate
interactive search sessions on Mars using data from the TREC Sessions
Track. In both cases, our work examines what researchers might call
``reasonable baselines''. We readily concede that we do not propose
any novel retrieval techniques {\it per se}---the value of our work
lies in formulating a novel problem and laying out the groundwork for
future explorations. As such, we hope that our contribution is
evaluated in terms of the vision it provides for future research on
interplanetary information retrieval.

\section{Background and Related Work}

The problem of searching from Mars is intended to be aspirational as
well as inspirational. Even if one remains unconvinced about
interplanetary colonization in the short term, our work remains
relevant in the same sense that zombie apocalypse preparations
advocated by the Centers for Disease Control are
instructive.\footnote{\url{http://www.cdc.gov/phpr/zombies.htm}} Like that
effort, theoretical considerations about unlikely scenarios can lead
to insights with more immediate impact. In fact, search from Mars can
be thought of as a specific instantiation of what Teevan et
al.~\cite{Teevan_etal_HCIR2013} call ``slow search'', which aims to
relax latency requirements for a potentially higher-quality search
experience. Slow search explores latencies on the order of minutes to
hours, which is similar to speed of light propagation delay to
Mars. There is substantial precedent for our work, as we discuss
below.

Technologies developed for search on Mars have potential applications
closer to home in improving search from remote areas on Earth such as
Easter Island, where only satellite Internet is available, and the
Canadian Arctic, where Internet access remains prohibitively slow and
expensive. Our work builds on previous efforts to enhance Internet
access in developing regions such as rural India, where connectivity
is poor and intermittent. Thies et al.~\cite{Thies_etal_WWW2002} explored web search
over email, an interaction model that is not unlike searching from
Mars. Chen et al.~\cite{ChenJay_etal_WWW2009} specifically tackle the problem of search
over intermittent connections, attempting to optimize the amount of
interaction that a single round of downloading can
enable. Intermittent connections can be modeled as high latency, which
makes the problem quite similar to ours---and indeed Chen et al.\ use
some of the query expansion and pre-fetching techniques we explore
here.

In this work, we assume that a functional interplanetary Internet
already exists, and that the only problem we need to overcome is
latency at the application layer. This is not an unrealistic
assumption as other researchers have been exploring high-latency network
links in the context of what is known as delay-tolerant networking
(see, for example, IETF RFC
4838\footnote{\url{https://tools.ietf.org/html/rfc4838}}) and NASA has
already begun experimental deployments on the International Space
Station.\footnote{\url{http://www.nasa.gov/mission_pages/station/research/experiments/730.html}}
Once again, there are many similarities between building interplanetary
connectivity and enhancing connectivity in developing
regions. Examples of the latter include DakNet~\cite{Pentland_etal_2004}, deploying wifi
access points on buses to provide intermittent connectivity to users
along their routes and the work of Seth et al.~\cite{Seth_etal_2006} to ferry data
using mechanical backhaul (i.e., sneakernet)---which isn't very
different from our proposal to put a cache of the web on a Mars-bound
rocket (more details later).

Even if one accepts the premise of Mars colonization, there may remain
skepticism about the importance of providing web search. While
challenges such as sustaining life, finding appropriate shelter, and
extracting energy are no doubt paramount, the psychological health of
Martian colonists is important also. As the web has become an integral
part of our daily lives, we believe that replicating the experience of
the web on Mars is an integral element of maintaining psychological
well-being. The HI-SEAS (Hawaii Space Exploration Analog and
Simulation) missions and other previous efforts, which attempt to
simulate long-duration habitation on Mars, are a recognition that
keeping colonists sane is just as important as keeping them alive.

Having accepted the premise of searching from Mars, let us next flesh
out some of the constraints in more detail. There exist technologies
built around laser-based communication where it is possible to achieve
good bandwidth between Earth and Mars. The Lunar Laser Communications
Demonstration achieved a 622-Mbps downlink and a 20-Mbps uplink
between the Earth and the
Moon,\footnote{\url{http://llcd.gsfc.nasa.gov/}}
so something like this to Mars is technologically feasible.
More bandwidth can be achieved by
building more satellites, so we can probably count on ``reasonable''
bandwidth between Earth and Mars. The other important factor is
physical transit time from Earth to Mars on rockets, which we can use
as a vehicle for physically delivering a cache of data (i.e., an
interplanetary sneakernet). Missions to Mars have taken between 150
and 300 days over the past half
century,\footnote{\url{http://www.universetoday.com/14841/how-longdoes-it-take-to-get-to-mars/}}
and without getting into details about orbital mechanics (tradeoffs
between transit time, fuel efficiency, and the prevalence of suitable
launch windows), it suffices to say physical transport between the two
planets will be on the order of months with current rocket
technology. Physical transport time defines a ``cache invalidation''
problem---as whatever data we put on a rocket runs the risk of
becoming stale before it arrives on Mars.

This paper builds on two previous papers that have tackled the search
from Mars problem. The first is an unrefereed magazine column that to
our knowledge is the first articulation of the search from Mars
problem~\cite{Lin_etal_IEEE2016a}. That article articulates the vision, but lacks
technical depth.
The second is a short paper~\cite{Clarke_etal_ICTIR2016} that empirically examines the high-recall
problem, which provide the basis of a more detailed study we describe
in Section~\ref{recallSec}.

\section{Spacetime Tradeoffs}
\label{tradeSec}

Achievable response times for searching on Mars requires a tradeoff between
latency and bandwidth.
If the available bandwidth between Earth and Mars is very large,
with few restrictions on usage,
searching on Mars need be little different than searching on Earth.
Mars would maintain a snapshot of the Earth-based portion of the web on local
servers (initially delivered by sneakernet), continuously updating it with the help of Earth-based crawlers.
Although this cache would still (unavoidably) be 4 to 24 minutes behind Earth,
a searcher on Mars would experience no lag.
Of course, if a search on Mars leads the searcher to an Earth-based
transactional site, or to other dynamic content, that site will still be
subjected to response time delays unless it too provides accommodations for
extraterrestrial users, an issue we leave for future work.

Unfortunately, maintaining a snapshot of the Earth-based web means that much
of the transferred data will go unseen and unused, at least until the colony
gains a sizeable population.
Furthermore, although details regarding communications technology are far from finalized,
we imagine that bandwidth will be limited and must be used parsimoniously.
While some bandwidth might be usable for speculative pre-fetching and
caching, potentially wasteful usage must be justifiable by potential
benefits to the colonists.

At the other extreme in this design space,
if available bandwidth between Earth and Mars is very limited,
with usage restricted to the most critical purposes,
we can do little to improve searching on Mars.
Any kind of speculative pre-fetching or caching would waste critical resources.
Under these circumstances, our colonists must tolerate the lag,
along with the other restrictions of pioneer life.

Since bandwidth limitations are unknown at present,
we quantify tradeoffs in terms of two measurable values, both independent of bandwidth:
\begin{list}{\labelitemi}{\leftmargin=1em}
\setlength{\itemsep}{-2pt}
  \item[1.]
    It takes a Martian longer to perform an online task,
    relative to the time required on Earth.
    She requires additional time either because she has to wait longer
    for an interaction to happen, or because she does extra work to
    complete her task.
    For example, while waiting for a search result the Martian might
    work on some unrelated task, or she might continue to peruse the results
    of a previous search while she waits for the new results to arrive.
    In this paper, we do not make a strong distinction between waiting and
    extra work, since both are wasted effort.
  \item[2.]
    We can send more data to Mars, relative to the amount of data we would
    send to a user's interaction device on Earth.
    For example, we might send extra search results, web pages, etc.\ that
    the user might never actually view. Two possible techniques are to pre-fetch results and to cache a portion of the web on Mars.
\end{list}

We can express the first value as an ``effort ratio'', $E$, where
user effort might be measured in task completion time,
or in some proxy, such as number of web pages viewed:
\begin{equation*}
E~=~\frac{
  \mbox{user effort required to complete task on Mars}
}{
  \mbox{user effort required to complete task on Earth}
}
\end{equation*}

We express the second value as a ``data ratio'', $D$, where data volume
might be measured in bytes,
or in some proxy, such as number of web pages transferred:
\begin{equation*}
D~=~\frac{
  \mbox{data transferred to complete task on Mars}
}{
  \mbox{data transferred to interaction device on Earth}
}
\end{equation*}

For interactive web search, there is a tradeoff between $D$ and $E$.
If we perform no pre-fetching or caching on Mars,
using the search engine exactly as on Earth, we have $D = 1$ but $E$
is maximized.
If we continuously send a full web crawl to Mars, we get $E = 1$,
but $D$ is maximized.
On Earth, $D = 1$ and $E = 1$ by definition.

On Mars, we trade off one against the other.
While $E$ is determined largely by the distance between the two planets
and the number of roundtrip delays required to complete a task,
$D$ may be arbitrarily large, even when little interaction is required.
For example, even pre-fetching pages linked from a SERP
(see Section~\ref{serpSec}) increases the number of pages transferred by
roughly a factor of ten, even if the user only clicks on a few (or no) results.

\begin{figure}[t]
\centering\includegraphics[width=0.99\linewidth]{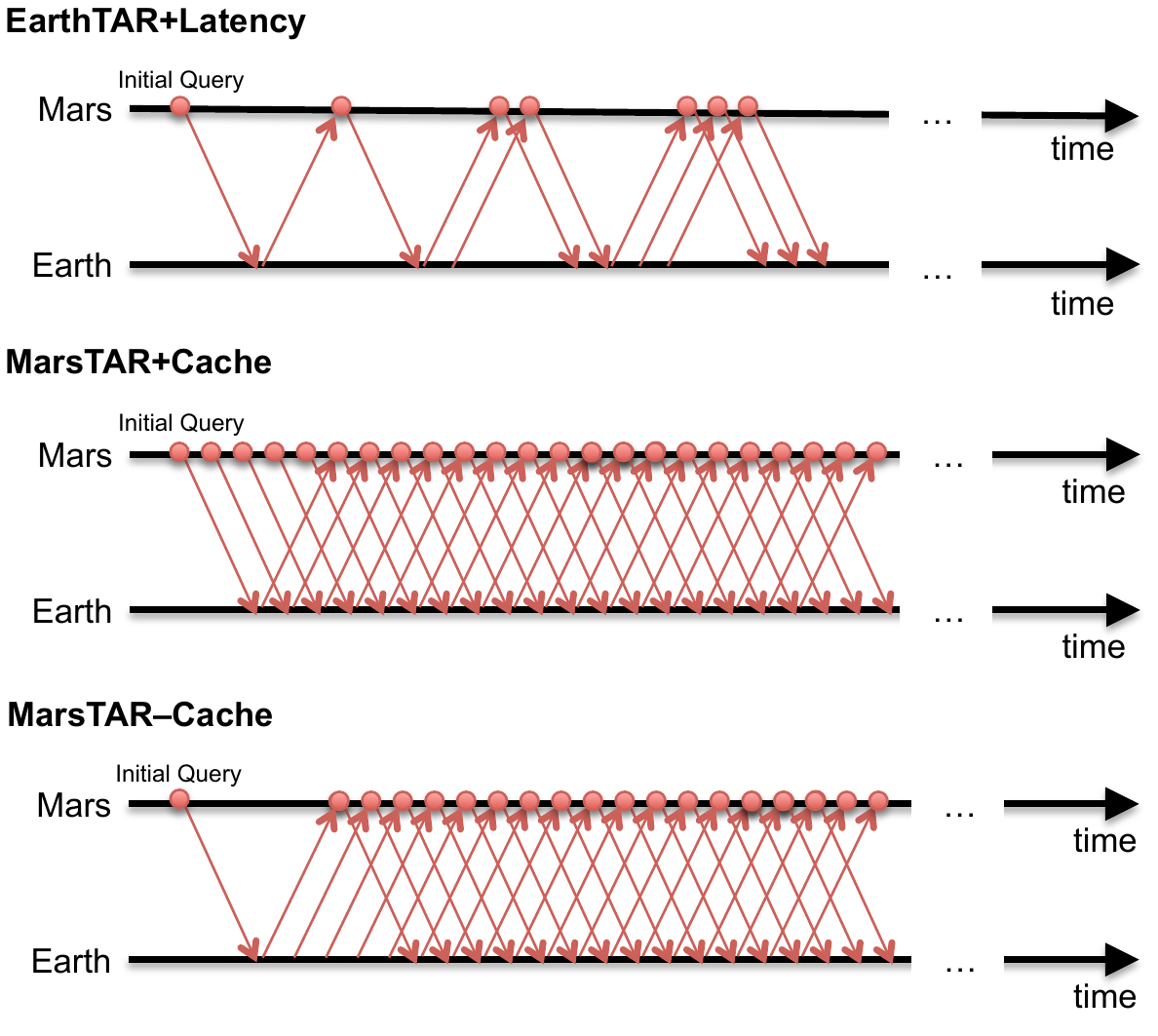}
\vspace{-0.6cm}
\caption{Illustration of various AutoTAR on Mars scenarios. Circles
  indicate relevance judgments.}\label{figure:scenarios}
\vspace{-0.3cm}
\end{figure}

\begin{figure*}
\centering
\includegraphics[width=0.49\linewidth]{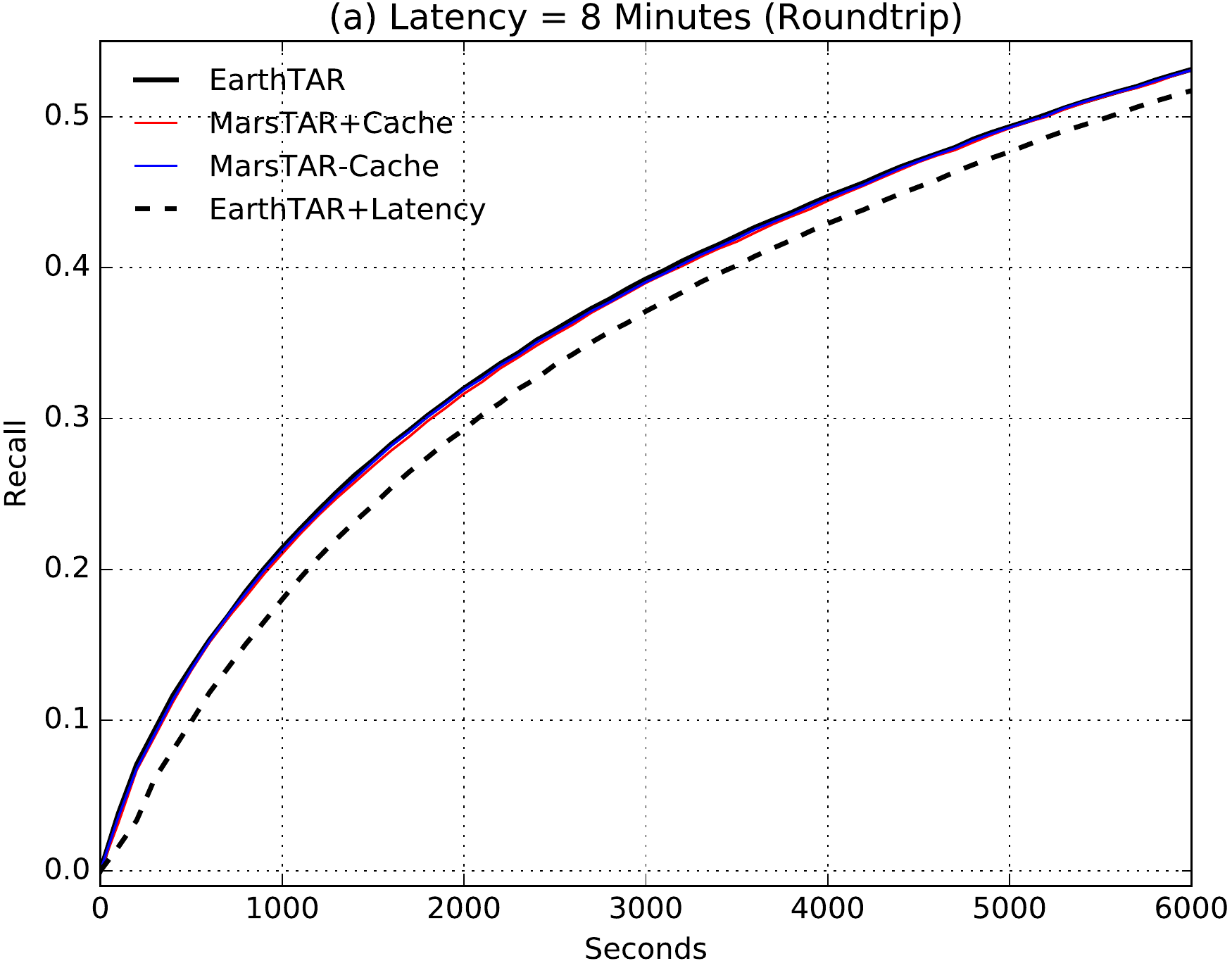}
\includegraphics[width=0.49\linewidth]{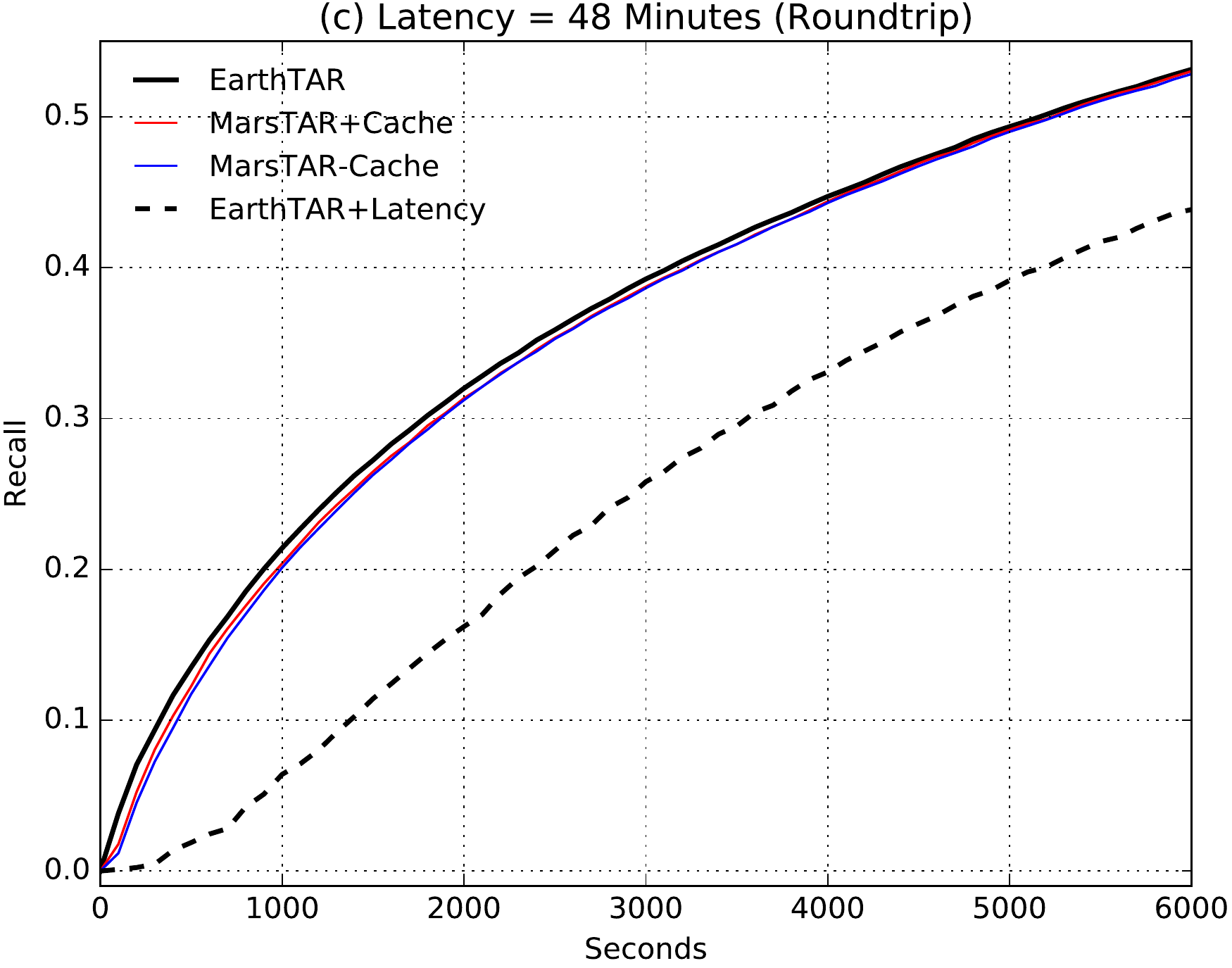}
\vspace{-0.3cm}
\caption{A comparison of the four formulations of AutoTAR with 8 and 48 minute roundtrip delays.}
\label{fig:tr-replication}
\vspace{-0.3cm}
\end{figure*}

\section{Case Study 1: Total Recall}
\label{recallSec}

As an example of the tradeoffs discussed in the previous section,
we revisit a previous study
that considered the problem of high-recall tasks, e.g., scientific surveys, in the context
of a permanent Martian colony~\cite{Clarke_etal_ICTIR2016}.
In this context, the Martian searcher aims to find as much relevant material as 
possible while minimizing the amount of non-relevant material consumed.

The previous study examined this task using Cormack and Grossman's AutoTAR protocol~\cite{AutoTAR,Cormack2016},
which uses continuous active learning~\cite{Cormack2014,Cormack2015} to iteratively train a classifier based on
searcher feedback collected (for computational efficiency) in batches. 
We omit the details of the underlying algorithm for brevity and encourage interested readers
to review the literature.

Four formulations of the AutoTAR protocol were purposed in Clarke et al~\cite{Clarke_etal_ICTIR2016}.
As we replicate this previous study using more realistic measurements of time, we provide
a brief description of these formulations (See Figure~\ref{figure:scenarios}):
\begin{list}{\labelitemi}{\leftmargin=1em}
\setlength{\itemsep}{-2pt}
\item {\bf EarthTAR}: The baseline result of running AutoTAR on Earth---the upper bound of performance.
\item {\bf EarthTAR$+$Latency}: The result of running AutoTAR from Mars without any attempts to hide latency, i.e., the searcher waits between batches of new documents from Earth to assess.
\item {\bf MarsTAR$+$Cache}: Two versions of AutoTAR are running, one on Earth and one on Mars. The Martian AutoTAR begins by running on a cache that has been shipped to Mars and is incrementally added to as Earth identifies new potentially-relevant documents that are shipped over to Mars. Earth runs its own version of AutoTAR on the entire collection and is trained by Martian assessments (received after a delay).
\item {\bf MarsTAR$-$Cache}: As above, except there is no pre-existing Martian cache. Thus, after the initial query, the Martian must wait for a roundtrip latency before she can begin assessing documents---these documents are always the ones sent from Earth.
\end{list}

The last three formulations were compared to the EarthTAR upper bound
to examine the impact of various round\-trip latencies.
As before, we used the Reuters Collection Volume 1 (RCV1)~\cite{lewis2004}, which comes with a fully labeled training and test set split over 103 topics. 
The training portion, the chronologically first $\sim$24,000 documents of the collection, was used as the Martian cache where applicable.
In previous work, we made a simplifying assumption, that the time to judge a document was one light minute, i.e.,
in four light-minutes of latency between Earth and Mars, the searcher could judge four documents.
Such an assumption makes for easy computation but is not necessarily realistic as it may take seconds to judge a document
but minutes to transmit a document. 
To more accurately model the real scenario, we use Smucker and Clarke's model~\cite{Smucker_Clarke_SIGIR_2012} of reading speed to calculate time to read and
judge a document for our Martian searchers. 
Their model predicts that for a document with {\it l} words it would take $0.018\textit{l}+7.8$ seconds
to read the document.

The assumption that reading time was equivalent to transmission time meant that our previous work
used arbitrary latencies between Earth and Mars, ranging from 5 units of time to 500 units of time. 
In our replication, we use roundtrip latencies of 8 and 48 minutes---these function as minimum and maximum delays that we might expect.

Figure~\ref{fig:tr-replication} plots the recall achieved as a function of the amount of time spent searching, including any
perceived latency, for the two levels of roundtrip latency.
These results agree with those reported previously but provide a more accurate picture of what might be conceivably experienced
on Mars. 
It is apparent that a small Martian cache is sufficient to achieve comparable performance with Earth. 
Even with no cache, Mars is quickly able to ``catch up'' to Earth-like effectiveness.  
Furthermore, we note that in the context of high-recall retrieval, there is no wasted transmission.
That is, to be sure all relevant material is identified, a searcher must exhaustively examine the entire corpus,
regardless of the underlying retrieval protocol, rather than traditional web search where a searcher may examine only one or two documents returned by a SERP.
It is worth noting that at any given point in time, the MarsTAR solution may have resulted in additional non-relevant documents 
being received by Mars when compared to EarthTAR.
The plots show, however, that such a discrepancy appears to be relatively small. 

\begin{figure*}[t]
\begin{small}
\begin{verbatim}
  ...
  </interaction>
  <interaction num="2" starttime="123.2863" type="reformulate">
     <query>Indian miss universe political issues.</query>
     <results>
        <result rank="1">
           <url>www.adpunch.org/entry/lara-dutta-wrapped-in-leaves/</url>
           <clueweb12id>clueweb12-1304wb-65-15002</clueweb12id>
           <title>Lara Dutta, wrapped in leaves</title>
           <snippet> ... Lara Dutta? No? Lara Dutta is an Indian actress, UNFPA Goodwill Ambassador and former
                     Miss Universe 2000, says Wikipedia.lara dutta b 080710 ... </snippet>
        </result>
        <result rank="2">
           <url>http://www.missuniverse.com</url>
 ...
\end{verbatim}
\end{small}
\vspace{-0.5cm}
\caption{Example of TREC 2014 Session Track log data.}
\label{logFig}
\vspace{-0.25cm}
\end{figure*}

\section{Case Study 2: Search Sessions}

In our second case study we use an existing query log to examine the
impact of searching from Mars and to explore pre-fetching and
caching techniques for remediating Earth-Mars latencies.
For each search session in our log, we plot the number of pages transferred
to the user's interaction device against the total time of the session.
For searching on Earth, these numbers come directly from the log,
since these sessions actually did take place on Earth.
For searching on Mars, we add an appropriate delay for each interaction with
the search engine.
We look particularly at the two extreme cases,
an 8-minute delay when the planets are at their closest,
and a 48-minute delay when the planets are farthest apart.
For this simple simulation, we assume the user waits
(or works on some other task) during each interaction cycle.

Academic research into web search is hampered by the paucity
of query log data, particularly data for complete search sessions.
To address this need, the TREC Session Track created test collections
and evaluation measures to study search across multi-query
sessions~\cite{session10, session11, session12, session13, session14}, as opposed to single-shot queries.
As part of this effort, the track organizers gave specific search tasks
to recruited users, and recorded queries, clicks, dwell times, and
other information as the users conducted these search tasks.
The track ran for five years (TREC 2010-2014).
We used TREC 2014 data for our experiments~\cite{session14}.

For TREC 2014, track organizers recruited users through Amazon Mechanical Turk,
recording 1,257 unique sessions, comprising 4,680 queries and 1,685 clicks.
Users conducted searches with the
Indri search engine 
over the ClueWeb12 collection,
a crawl of 733 million pages from the general web gathered in early 2012.\footnote{\url{http://lemurproject.org/clueweb12/}}
While the size of this collection is modest by commercial standards,
and the size of the log is dwarfed by a few milliseconds of commercial search,
it has proven to be a valuable resource for understanding user behavior across
sessions~\cite{ysw16}.
For illustrative purposes, a sample of the log is shown in Figure~\ref{logFig}.


\subsection{Baselines}

\begin{figure}
\includegraphics[width=\columnwidth]{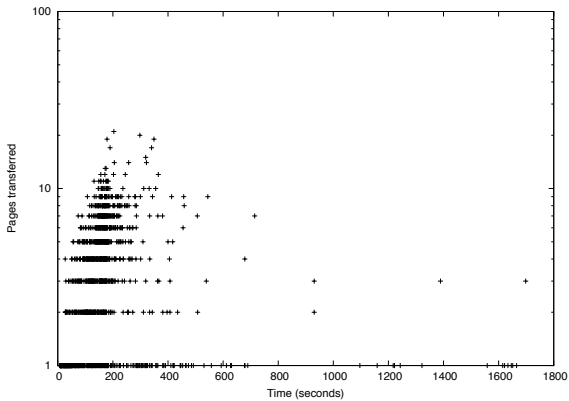}
\vspace{-1.0cm}
\caption{Search Sessions on Earth:\ Each point indicates the pages transferred and
  total time for a single session.}
\label{EarthPlot}
\vspace{-0.25cm}
\end{figure}

Earth-based interactions are taken directly from the log,
which was naturally recorded on Earth.
Figure~\ref{EarthPlot} plots sessions,
with each point representing a single session.
Session duration is plotted on the $x$ axis and the number of pages transferred
is plotted on the $y$ axis.
For the purposes of counting pages transferred,
a SERP counts as a single page and a click counts as a single page.

Points falling along the $x$ axis represent searches where the user issued
only a single query and did not click on any results.
Most sessions take under ten minutes, with a few taking nearly a half hour.
In these logs, session duration is assumed to start when the user begins
to consider the search problem, and not when the first query is issued.
We retain this approach in our experiments.

\begin{figure*}
  \begin{minipage}{0.5\textwidth}
    \includegraphics[width=\columnwidth]{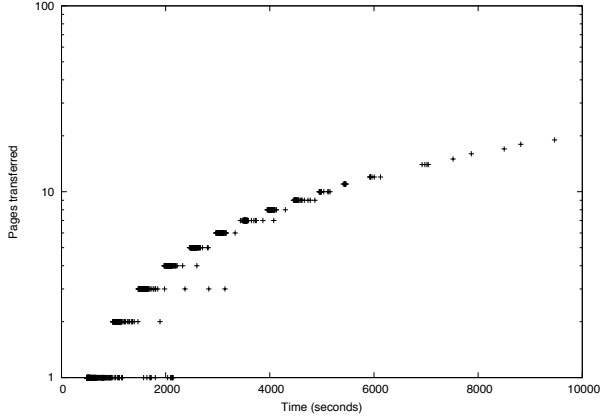}
    \vspace{-1.0cm}
    \subcaption{8 minute response time}
  \end{minipage}
  \begin{minipage}{0.5\textwidth}
    \includegraphics[width=\columnwidth]{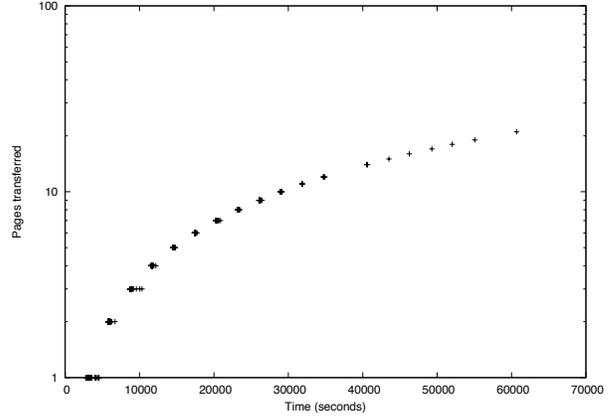}
    \vspace{-1.0cm}
    \subcaption{48 minute response time}
  \end{minipage}
\vspace{-0.2cm}
\caption{Search Sessions on Mars:\ Each point represents a single session with no pre-fetching or caching.}
\label{MarsPlot}
\vspace{-0.5cm}
\end{figure*}

As the simplest simulation of searching from Mars, we can replay the session logs,
assuming the user waits (or does other work) after each query and click,
while the request is sent to Earth and the response is returned to Mars.
Figure~\ref{MarsPlot} plots sessions under minimum and maximum delay times.
In these plots, delays clearly dominate interaction times,
especially with a worst-case 48~minute delay.
These simulations do include some very basic caching.
If a query is issued multiple times or if a page is clicked multiple times,
we assume the result is fetched only once.

\begin{table}[t]
\begin{tabular}{lc|cccc}
         &       & Average & Average     & Effort & Data \\
         & Lag   & time    & pages       & ratio  & ratio\\
Location & (min) & (sec)   & transferred & ($E$)  & ($D$)\\
\hline
Earth & 0  &  172.323  &  3.940  &  1.000  &  1.000 \\
Mars & 8  &  2046.118  &  3.904  &  15.334  &  0.995 \\
Mars & 48  &  11415.092  &  3.904  &  87.005  &  0.995 \\
\end{tabular}
\caption{Average performance for Earth-based and Mars-based sessions
  under various delay scenarios, with no pre-fetching or caching.}
\label{baselineTable}
\vspace{-0.4cm}
\end{table}

Table~\ref{baselineTable}
shows average transfers and average session duration for various scenarios,
along with effort ratios ($E$) and data ratios ($D$) as defined in
Section~\ref{tradeSec}.
Average effort ratio ($E$) essentially grows linearly with roundtrip time,
i.e., the lag seen by the user.
Data ratio ($D$) actually drops slightly, since we do not assume caching
in an Earth-based browser.
If we had, 
we would have $D = 1$ in all cases.

\subsection{Pre-fetching}
\label{serpSec}

How might we begin to hide latencies associated with searching from Mars?
After the initial query in a session, we might attempt to predict the
user's needs and pre-fetch pages required for the remainder of the session,
potentially reducing $E$ at the cost of an increase in $D$.
We try three simple approaches:\ pre-fetching pages linked directly from SERPs (up to ten),
pre-fetching additional related pages (perhaps several thousand)
along with the pages linked from SERPs, and
expanding with query suggestions and returning associated SERP pages.


\begin{figure*}[t]
  \begin{minipage}{0.5\textwidth}
    \includegraphics[width=\columnwidth]{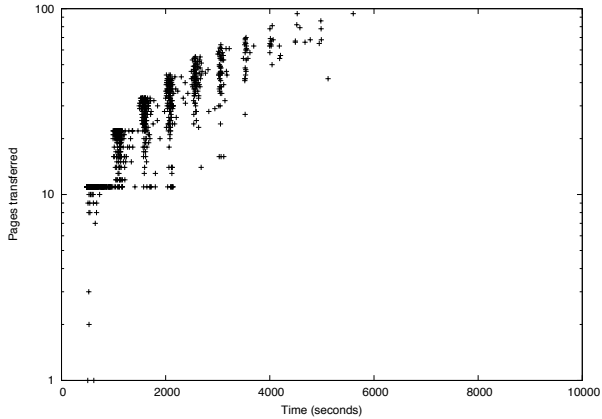}
    \vspace{-1.0cm}
    \subcaption{8 minute response time}
  \end{minipage}
  \begin{minipage}{0.5\textwidth}
    \includegraphics[width=\columnwidth]{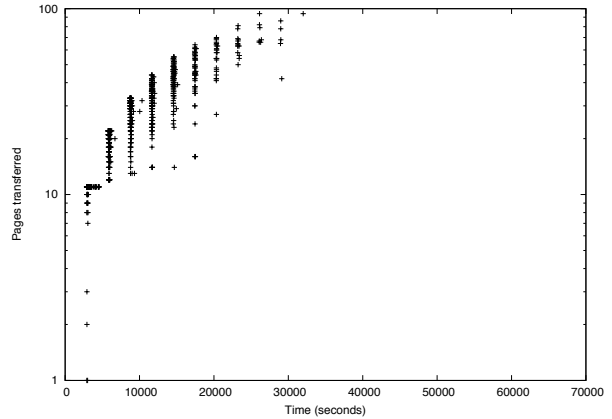}
    \vspace{-1.0cm}
    \subcaption{48 minute response time}
  \end{minipage}
\vspace{-0.2cm}
\caption{Search Sessions on Mars:\ Each point represents a single session with SERP pre-fetching.}
\label{SERPPlot}
\vspace{-0.2cm}
\end{figure*}

\smallskip \noindent {\bf SERP Pre-fetching.}
As our first attempt at pre-fetching to reduce $E$, we pre-fetch result pages
linked from SERPs, under the assumption that the user will click on at
least some of them.
Indeed, pre-fetching of result pages is so obviously sensible that we
cannot imagine supporting search on Mars without at least this form of 
pre-fetching, unless bandwidth limitations are extremely severe.
Here, we pre-fetch only the pages directly linked from SERPs in the logs,
but we might imagine going further, perhaps by loading more of the linked site
or by pre-fetching deeper in the result list.

\begin{table}
\begin{tabular}{lc|cccc}
         &       & Average & Average     & Effort & Data \\
         & Lag   & time    & pages       & ratio  & ratio\\
Location & (min) & (sec)   & transferred & ($E$)  & ($D$)\\
\hline
Earth & 0  &  172.323  &  3.940  &  1.000  &  1.000 \\
Mars & 8  &  1436.667  &  23.593  &  11.263  &  7.477 \\
Mars & 48  &  7758.385  &  23.593  &  62.578  &  7.477 \\
\end{tabular}
\caption{Average performance for Earth-based and Mars-based sessions
  under various delay scenarios, with SERP pre-fetching.}
\vspace{-0.4cm}
\label{SERPTable}
\end{table}

Figure~\ref{SERPPlot} plots individual sessions using SERP pre-fetching,
while Table~\ref{SERPTable} shows average effort and data ratios.
While most SERPS in our query log link to ten pages, $D$ is less than 10
due to caching effects.
$E$ increases linearly with lag, but values are at 25\% or more below those
in Table~\ref{baselineTable}. That is, we save around 25\% effort at the cost
of transferring around around seven times more data than necessary.

\smallskip \noindent {\bf Topical Pre-fetching.}
A simple way to extend SERP pre-fetching would be to go deeper in the ranked list,
perhaps uploading a large set of topically related pages in response to each
query, along with the pages linked directly from the SERP.
Even when a Martian user issues a further query,
requiring another roundtrip delay,
these pages would allow the user to further explore their topic while awaiting
additional results from Earth.
If large numbers of related pages are uploaded,
query reformulations can also be issued against these local pages,
perhaps allowing the Martian to satisfy their information need without waiting.

To explore the potential for topical pre-fetching,
we indexed the ClueWeb12 collection using the Lucene search engine.
As queries appear in the log, we execute them against this index using BM25.
We then assume that the top $k$ documents are sent to Mars,
along with documents linked from the SERP.
Since we are using different search tools (and ranking algorithm) from those used by the TREC Session
Track, not all pages from the SERPs appear in our top $k$.
In reality, of course, the SERP documents would be the top-ranked subset of
our top $k$, so that exactly $k$ documents would be transferred to Mars.

We consider hits on these uploaded documents,
where we count as a hit any topically pre-fetched document that later appears
in a SERP from the same session.
Having these pages already on Mars potentially allows the Martian user to
access them without having to wait for the SERP in which they first appear.
With $k=1000$ we achieve a hit ratio of over 21\%;
with $k=2000$ we achieve a hit ratio of over 27\%.
These hit ratios should translate into substantial reductions in $E$,
although a reasonable estimate requires many assumptions about user behavior,
which we avoid in this paper.
Unfortunately, this potential improvement to $E$ comes at a cost in $D$,
as compared to SERP pre-fetching alone, since $D$ is approximately equal to $k$.
That is, we marginally improve effort at a great cost in transferring data
that is never used.

\smallskip \noindent {\bf Query Suggestions.}
Most commercial search engines suggest query reformulations and extensions to
help guide their users through search sessions.
We might take advantage of these suggestions by executing them on behalf of
the Martian user, uploading the results and their linked pages,
along with the main SERP and its linked pages.
If a reformulation by a Martian user matches a query suggestion,
we completely avoid a query-response cycle to Earth,
Even if the Martian makes an unanticipated reformulation,
the additional uploaded information might allow her to continue working while
waiting for a response from Earth.

To explore the potential of this idea,
we submitted queries from our log to the Bing Autosuggest API\footnote{\url{https://www.microsoft.com/cognitive-services/en-us/bing-autosuggest-api}}
and compared suggestions to queries appearing later in the same session.
For 57~queries, a suggested query appeared verbatim later in the same session.
While this is less than 2\% of all possible queries,
it is clear that the idea has some potential,
perhaps by going deeper in the suggestion list or by extracting related terms
from suggested queries.
While some of the suggestions are spelling corrections or simple morphological
variants, some are more complex, e.g., ``uglai recipe'' for ``kenya cooking''.

Combining our various query pre-fetching ideas may provide a reasonable
overall solution.
When a query is received from Mars,
we might imagine expanding it with terms from query suggestions,
and through other expansion methods,
we could generate a large set of related documents
to transmit to Mars for re-ranking and exploration by the
Martian user.
We might even stream documents continuously,
similar to Section~\ref{recallSec},
adjusting the stream on the basis of queries and other feedback from Mars.
We leave the investigation of such ideas for future work.

\subsection{Caching}

As an alternative or in addition to pre-fetching, we could minimize user effort by (partially) caching
a snapshot of the web on Mars (we discuss the possible logistics below).
If we maintain a partial snapshot on Mars, perhaps we could serve most of
the user traffic from that cache, or at the very least give the user some
preliminary results to work with while we are fetching full results.
But of course, much of the web consists of lower quality pages that
would rarely appear in a SERP, and would even more rarely receive a click---the question,
of course, is which parts of the web do we send over to Mars?
Caching will greatly increase $D$,
but if the pages are selected based on some type of static rank,
or ``page quality'', we may be able to reduce $E$.

The experiments in this paper used the ClueWeb12 crawl and TREC session logs.
Despite possible concerns about the realism and fidelity of the data, 
we nevertheless can still gain some insight regarding the value of caching.

For static ranking, we use the method of Cormack et al~\cite{csc11},
which has performed well on ClueWeb collections
(the source of the Waterloo spam scores that are widely used by academic researchers)
and has fast code available.
Static ranking is based on content only.
We trained over the ClueWeb09 collection---an earlier crawl gathered by
the same group at CMU in 2009---using as training labels TREC relevance
assessments created as part of various experiments using that collection.
More specifically, we trained on:\
1) all documents judged relevant for any purpose (e.g., for any query)
regardless of grade, which were taken as positive examples;
2) all documents assessed as spam, which were taken as negative examples; and
3) a random sample ($N = 3000$) of documents judged as non-relevant,
which was also taken as negative examples. The static ranker was then applied
to all pages in ClueWeb12. Note that training data is completely disjoint from this collection, and
so there is no ``information leakage'' from the session data.

\begin{figure}[t]
\vspace{-0.4cm}
\includegraphics[width=\columnwidth]{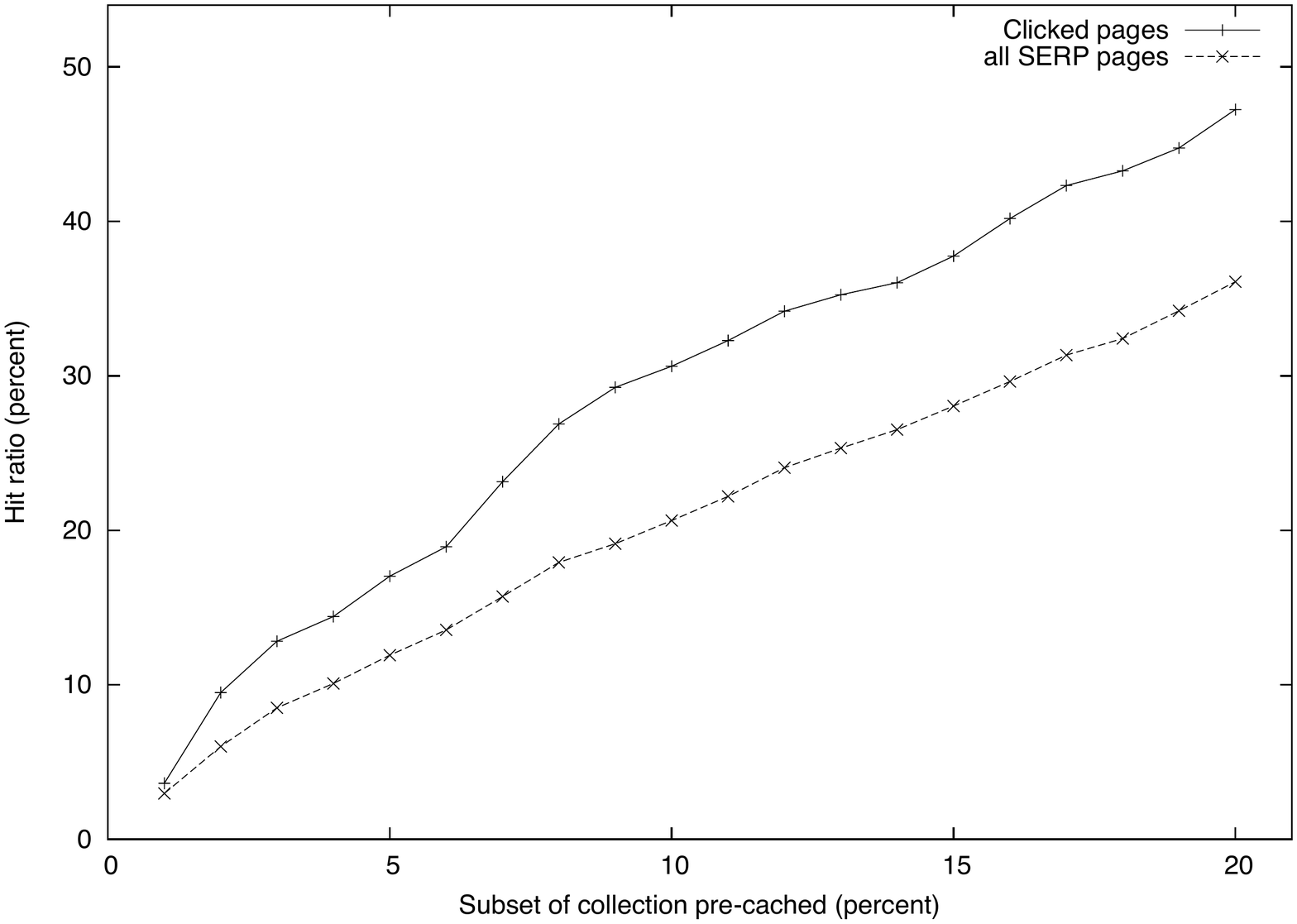}
\vspace{-1.0cm}
\caption{Cache hit ratios for clicked pages and for all SERP result pages.}
\label{caching}
\vspace{-0.4cm}
\end{figure}

\begin{figure*}[t]
  \begin{minipage}{0.5\textwidth}
    \includegraphics[width=\columnwidth]{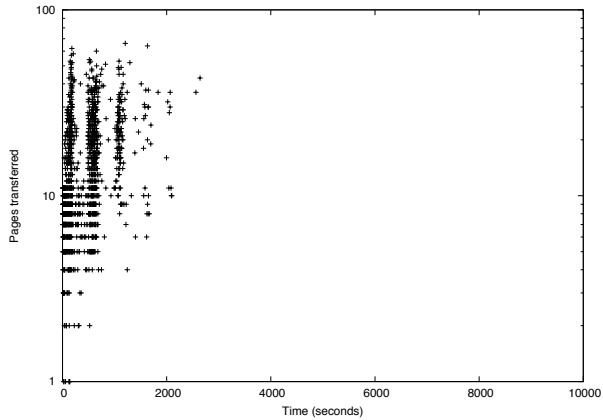}
    \vspace{-1.0cm}
    \subcaption{8 minute response time}
  \end{minipage}
  \begin{minipage}{0.5\textwidth}
    \includegraphics[width=\columnwidth]{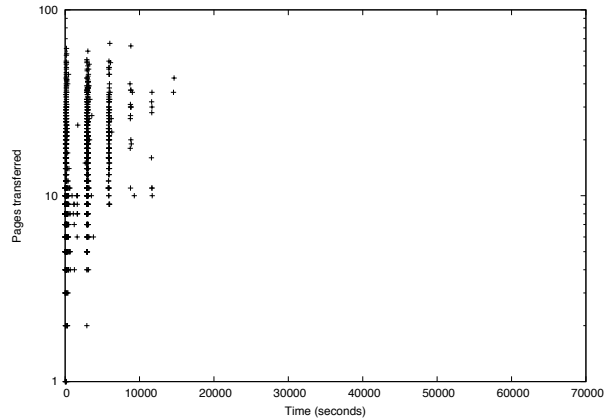}
    \vspace{-1.0cm}
    \subcaption{48 minute response time}
  \end{minipage}
\vspace{-0.2cm}
\caption{Search Sessions on Mars:\ Each point represents a single
  session with SERP pre-fetching and 20\% caching. The range of
  the $x$ axis is the same as Figures~\ref{MarsPlot}
  and~\ref{SERPPlot}. Pages transferred exclude pages in the cache.}
\vspace{-0.2cm}
\label{precachePlot}
\end{figure*}

Based on this static ranking, we might cache a fraction of available
pages on Mars.
Figure~\ref{caching} shows hit ratios for cached pages appearing
in the log, considering either all pages linked from SERPs or just pages
that were actually clicked.
Hit ratios are shown for various caching ratios between 1\% and 20\% of the
full collection.
The hit ratio for clicked pages is substantially higher than that for SERP
pages generally, helping to confirm the success of our static ranking.
By maintaining a 20\% snapshot on Mars, we can achieve a hit ratio for clicked
pages of nearly 50\%.

To simulate the impact of caching pages on Mars,
we require some assumptions about user behavior in addition to the actual
behavior captured in the log.
Each session starts as usual, with the user issuing the query appearing 
in the log.
The query is sent to Earth,
which follows the SERP pre-fetching approach in Section~\ref{serpSec},
returning the SERP itself and all pages linked from the SERP that are not
already on Mars.
Meanwhile, the query is also sent to the local cache,
which we assume returns a SERP covering the pages in the cache.
The user interacts with this local SERP until the log shows she would have
clicked on a result not present in the local cache.
At that point our simulated user waits for the full Earth-generated
SERP before proceeding.

If the user issues further queries, we follow the same process, sending the
query to Earth and allowing the user to interact locally without delay.
Delay occurs only when the log shows a click on a result not present in the
local cache.
While the Mars-based user would not actually be able to click on non-local
results, since they would not appear on the locally generated SERP,
we take the click as a signal of dissatisfaction with the local result.
Since we have no way of knowing how the real user would have proceeded,
waiting for the Earth-based results provides a worst-case assumption that
places an upper bound on $E$.

Figure~\ref{precachePlot} plots the results of this simulation with 20\% caching.
Here the $y$ axis shows only pages transferred beyond those already cached.
Compared with Figures~\ref{MarsPlot} and~\ref{SERPPlot},
overall session times are substantially reduced,
although they are still well above the Earth-based times in Figure~\ref{EarthPlot}.
Average performance appears in Table~\ref{precacheTable};
average pages transferred and data ratios exclude the
157~million cached pages.

\begin{table}[t]
\begin{tabular}{lc|cccc}
         &       & Average & Average     & Effort & Data \\
         & Lag   & time    & pages       & ratio  & ratio\\
Location & (min) & (sec)   & transferred & ($E$)  & ($D$)\\
\hline
Earth & 0  &  172.323  &  3.940  &  1.000  &  1.000 \\
Mars & 8  &  445.918  &  16.351  &  2.812  &  5.148 \\
Mars & 48  &  1936.334  &  16.351  &  12.561  &  5.148 \\
\end{tabular}
\vspace{-0.25cm}
\caption{
  Average performance for Earth-based and Mars-based sessions
  under various delay scenarios, with SERP pre-fetching and 20\% caching.
  Average pages transferred and data ratios exclude the 157~million cached
  pages.
}
\vspace{-0.4cm}
\label{precacheTable}
\end{table}

How might we actually cache a snapshot of the web on Mars? While in
our simulations, 20\% of the collection represents a ``mere'' 157
million pages, 20\% of {\it the entire web} remains a substantial,
even daunting amount of data. The most practical approach is to
physically transport the cached data on cargo rockets (i.e., a
sneakernet). The problem, of course, is the transit time:\ many of the
pages will already have changed by the time the cache arrives at
Mars. Physical transport of data needs to be accompanied by updates
sent from Earth---which of course consumes valuable bandwidth. Without
updates, searchers on Mars would be interacting with a copy of the web
that is several months old.

The combination of sneakernet and incremental updates frames an
optimization problem that commercial web search engines are equipped
to solve. Today, they must decide what and how frequently to recrawl
existing content, and as a result have detailed historic data
indicating which pages are ``stable'' and which pages change
rapidly. With this information, it is possible to trade off physical
data transport with bandwidth-consuming updates. Although we do not
have access to this information, it is a matter of engineering to
figure out the best solution. This is a solvable problem.

\section{Conclusions and Future Work}

In this paper, we provide a framework for evaluating search from Mars
as a tradeoff between ``effort'' (waiting for responses from Earth)
and ``data transfer'' (pre-fetching or caching data on Mars). The
contribution of our work is articulating this design space and
presenting two case studies that explore the effectiveness of baseline
techniques. Although we do not present any novel retrieval techniques,
tackling the problem must begin with ``obvious'' solutions. We find,
indeed that we can trade off effort with the amount of data transferred, with varying
degrees. These simple techniques set the groundwork for future
studies.

As we noted earlier, the problem of searching from Mars has analogs
closer to home. Instead of a cache we ship to Mars on a cargo rocket,
we might FedEx hard drives of web data to rural villages in the
developing world, where the village elders can plug these caches into
the central wifi access point. This shared access point can intercept
web searches with the local cache; usage log data can determine the
pages that arrive on next month's hard drive shipment. This scenario
parallels exactly search from Mars, and thus searching from Mars is
more than idle blue-sky speculation. Furthermore, the breakthroughs
that are needed---for example, better session models and models of
long-term user needs---stand to benefit web search in general.

Moving forward, we continue to consider the larger problem of
supporting access to web and social media services on Mars. In the
medium term, we hope to establish a full simulation of interaction
from Mars, allowing for the creation and testing of a full stack built
on top of appropriate high-latency networking technology. Our goal is
to create a fully tested and ready-to-go solution for use by future
colonists.
Exploration is perhaps one of the most innate
human drives, and since we're not rocket scientists, structural
engineers, geologists, or botanists, there are limited contributions we can make
that are on the ``critical path'' of sending humans to Mars. However,
as information retrieval researchers we can contribute information access solutions for humankind's next
great leap.

\balance

\end{document}